\documentclass{moriond}

\def\be{\begin{equation}}
\def\ee{\end{equation}}
\def\bea{\begin{eqnarray}}
\def\eea{\end{eqnarray}}

\begin{document}
\vspace*{4cm}
\title{BARYON ELECTROMAGNETIC FORM FACTORS AT BESIII}

\author{C. MORALES MORALES on behalf of the BESIII Collaboration}

\address{Helmholtz-Institut Mainz, 55099 Mainz, Germany}

\maketitle\abstracts{
The electromagnetic form factors are fundamental observables that parametrise the electric and magnetic structure of hadrons and provide a key to the strong interaction. 
The Beijing Spectrometer (BESIII) is located at the Beijng Electron Positron Collider (BEPCII), a symmetric $e^+e^-$ collider running at center-of-mass energies between 2.0 and 4.6 GeV. This wide energy range allows the measurement of electromagnetic form factors both from direct $e^+e^-$ annihilation and from initial-state radiation processes. 
Based on $157~\mathrm{pb}^{-1}$ of data collected at center-of-mass energies between 2.23 and 3.67 GeV, BESIII published results on the channel $e^+e^-\rightarrow p\bar{p}$. More recently, preliminary results from the analysis of the initial-state radiation process $e^+e^- \rightarrow p\bar{p}\gamma$ based on 7.41 $\mathrm{fb}^{-1}$ of data have also been released. Besides nucleons, all hyperons in the SU(3) spin 1/2 octet and spin 3/2 decuplet are energetically accessible at BESIII. Preliminary results from the $e^+e^-\rightarrow \Lambda\bar{\Lambda}$ channel and the $e^+e^- \rightarrow \Lambda_c \bar{\Lambda}_c$ channel from the charmed sector are already available. Furthermore, a world-leading data sample for precision measurements of baryon form factors was collected in  2015. This data will enable the measurement of baryon electromagnetic form factors with unprecedented accuracy.
}

\section{Introduction}
\begin{figure}[tb]
\begin{center}
\includegraphics*[width=85mm]{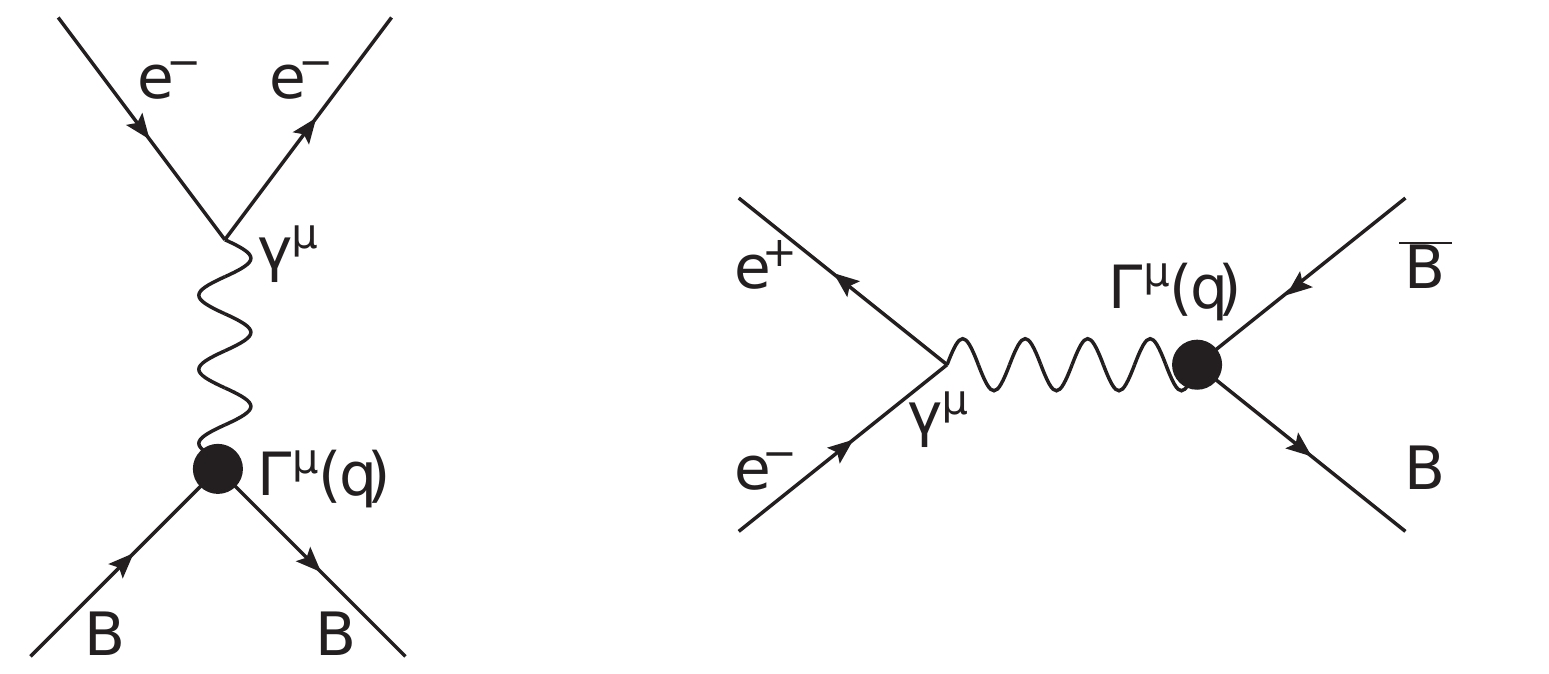}
\end{center}
\caption{Lowest-order Feynman diagrams for elastic electron-hadron scattering \mbox{$e^- B  \rightarrow e^- B$} (left), and for the annihilation process \mbox{$e^-e^+ \rightarrow B \overline{B}$} (right).}
\label{Feynman}
\end{figure}

Electromagnetic form factors (FFs) describe the modifications of the point-like photon-hadron vertex due to the internal structure of hadrons. Figure~\ref{Feynman} shows the lowest-order Feynman diagrams of the electron-hadron elastic scattering and the $e^+e^-$ annihilation into a pair of baryons. In the case of spin 1/2 baryons, the hadronic vertex is described by a non-constant matrix: 
\begin{equation}
\Gamma^\mu(q^2) = \gamma^\mu F_1(q^2) + \frac{i\sigma^{\mu\nu}q_\nu}{2m} F_2(q^2),
\end{equation}
where $F_1$ and $F_2$ are the so callled Dirac and Pauli FFs and $m$ is the mass of the corresponding baryon. The Dirac form factor, $F_1$, is related to the electric and the magnetic scattering from the baryon, while the Pauli form factor, $F_2$, is related to the additional scattering contribution arising from the anomalous magnetic moment of the baryon. Instead of $F_1$ and $F_2$, the use of the so-called Sachs FFs has become conventional:
\begin{equation}
G_E (q^2)= F_1(q^2) + \frac{q^2}{4m^2}F_2(q^2), \hspace{1.25cm}
G_M(q^2) = F_1(q^2) + F_2(q^2).
\end{equation}
The FFs are analytic functions of the momentum transfer $q^2$, 
and the normalization is done at $q^2 = 0$, with $F_1(0) = F_2(0) = 1$ and $G_E(0) = G_M(0)/\mu$ = 1, where $\mu$ is the magnetic moment of the baryon in units of the Bohr magneton.

Them FFs are real in the space-like region ($q^2<0$) and complex in the time-like region ($q^2>0$) above the production threshold for hadronic vector states ($q^2 > 4m_\pi^2$).  Space-like FFs can be measured in elastic scattering experiments (Fig.~\ref{Feynman} (left)), while time-like FFs are measured in annihilation processes \mbox{$e^+e^- \leftrightarrow B \overline{B}$} (Fig.1 (right)). Dispersion relations can be used to predict the FFs behavior in the unphysical region. 

Under the one-photon exchange approximation, 
the Born differential cross section of the process \mbox{$e^+e^- \rightarrow B\overline{B}$} in the $e^+e^-$ center-of-mass (c.m.) reads~\cite{Zichichi} 
\begin{equation}
\label{diff}
\frac{d\sigma^{\mathrm{Born}}(q^2,\theta_{B})}{d\Omega} = \frac{\alpha^2\beta C}{4q^2} \left [(1 + \mathrm{cos}^2\theta_{B}) |G_M (q^2)|^2 +  \frac{1}{\tau}\mathrm{sin}^2\theta_{B}|G_E(q^2)|^2 \right ],
\end{equation}
where $\theta_{B}$ is the polar angle of the baryon, $\tau = q^2/4m^2$, and $\beta = \sqrt{1-1/\tau}$. The Coulomb factor,  $C = y/(1-\mathrm{exp}(-y))$ with $y = \pi \alpha / \beta$, accounts for the electromagnetic $B \overline{B}$ interactions of point-like baryons~\cite{Tzara}, and is equal to 1 for neutral baryon pairs. 
The total cross section, integrated over the full solid angle is:
\begin{equation}
\sigma^{\mathrm{Born}}(q^2) = \frac{4\pi\alpha^2\beta C}{3q^2} \left[ |G_M(q^2)|^2 + \frac{1}{2\tau} |G_E(q^2)|^2 \right ]. 
\label{totalcs}
\end{equation}
An effective form factor (EFF) can be defined as
\begin{equation}
|G(q^2)|^2 =  \frac{2\tau |G_M(q^2)|^2 + |G_E(q^2)|^2}{2\tau +1} = \frac{\sigma^{\mathrm{Born}}(q^2) }{(1+\frac{1}{2\tau})(\frac{4\pi\alpha^2\beta C}{3q^2})} , 
\label{effectiveff}
\end{equation}
which is equivalent to $|G_M(q^2)|$ under the working hypothesis $|G_E(q^2)| = |G_M(q^2)|$. 
From the measurement of the differential cross section (Eq.~\ref{diff}) at a given $q^2$, it is possible to determine the ratio of the absolute value of the FFs, $R = |G_E/G_M|$. The precise knowledge of the normalization and the luminosity allows the separate independent of $|G_E|$ and $|G_M|$. 

Time-like electromegnetic form factors are complex functions of the momentum transfer with a relative phase $\Delta\Phi$. This phase induces polarisation effects in the final state. The polarisation of a baryon in a $e^+e^- \rightarrow B \bar{B}$ process, is perpendicular to the production plane spanned by the incoming beam and the outgoing baryon or antibaryon. The polarisation can be written as a function of the phase and the scattering angle of the baryon like~\cite{karin}: 
\begin{equation}
P_n = - \frac{\mathrm{sin}2\theta_B\sqrt{\tau}}{(1 + \mathrm{cos}^2\theta_B)+ \frac{R^2}{\tau}((1 - \mathrm{cos}^2\theta_B)}R\mathrm{sin}\Delta\theta,
\label{polarizacion}
\end{equation}
and the spin correlation of the outgoing baryon or antibaryon as:
\begin{equation}
C_{ml} = - \frac{\mathrm{sin}2\theta_B\sqrt{\tau}}{(1 + \mathrm{cos}^2\theta_B)+ \frac{R^2}{\tau}((1 - \mathrm{cos}^2\theta_B)}R\mathrm{cos}\Delta\theta.
\label{cml}
\end{equation}
Therefore, $\Delta\Phi$, and thus the modulus of the phase, can be extracted from the baryon polarisation. Hyperons have an advantage with respect to nucleons in the sense that they decay weakly. The interference of the decay amplitudes in the hyperon decay causes the daughter baryon to be emitted in the direction of the mother hyperon. Thus, the polarisation of the hyperon is accessible via the angular distribution of the daughter baryon and the spin corelations are accessible via the angular distributions of the daughter baryon and antibaryon. 

An alternative approach to measure hadronic cross sections at high luminosity $e^+e^-$ storage rings is the study of initial-state radiation (ISR) processes. 
The differential cross section of the ISR process $e^+e^- \rightarrow B \bar{B} \gamma$ is related to the cross section of the non-radiative process $e^+e^- \rightarrow B \bar{B}$ through
\begin{equation}
\frac{d^2\sigma^{\mathrm{ISR}}} {dq^2d\theta_{\gamma}} = \frac{1}{s} \cdot W(s,x,\theta_{\gamma}) \cdot \sigma^{\mathrm{Born}}(q^2),
\label{differential}
\end{equation}
where $x = 2E_\gamma/\sqrt{s} = 1 - q^2/s$, and $E_\gamma$ and $\theta_\gamma$ are the energy and the polar angle of the ISR photon in the $e^+e^-$ c.m., respectively. The radiator function, $W(s,x,\theta_{\gamma})$, describes the probability of the ISR photon emission~\cite{Radiator}.
The emission of an ISR photon leads to a reduction of the invariant mass of the final state, $q^2$, allowing for the measurement of $e^+e^- \rightarrow B \bar{B}$ cross sections over a continuous range of $q^2$ below the center-of-mass energy of the collider, $\sqrt{s}$.

\section{The Beijing Spectrometer at the Beijing Electron-Positron Collider}
The Beijing electron-positron collider, BEPCII, is a double ring symmetric collider running at $\sqrt{s}$ from 2.0 to 4.6 GeV. The design luminosity is $1 \times 10^{33}$~$\mathrm{cm}^{-2}\mathrm{s}^{-1}$ at a beam energy of 1.89 GeV and it was achieved in 2016. BESIII is a cylindrical detector which covers 93\% of the full solid angle. It consists of several sub-detectors. A small cell, hellium based ($60\%$ He, $40\%$ $\mathrm{C}_3\mathrm{H}_8$) Main Drift Chamber (MDC) which provides momentum measurements of charged particles with a resolution of 0.5$\%$ at 1 GeV/$c$ in a 1 T magnetic field. The energy loss measurement (dE/dx) provided by the MDC has a resolution better than 6$\%$.
A Time-of-Flight plastic scintillator (TOF) consisting of 5-cm-thick plastic scintillators with a time resolution of 80 ps in the barrel and 110 ps in the end caps of the detector. 
A CsI(Tl) Electro-Magnetic Calorimeter (EMC) consisting of 6240 crystals arranged in a cylindrical structure and two end caps used to measure the energies of electrons and photons. The energy resolution of the EMC for 1 GeV electrons and photons is 2.5$\%$ in the barrel and $5\%$ in the end caps. The position resolution of the EMC is 6 mm in the barrel and 9 mm in the end caps for 1 GeV electrons and photons. Finally, a Muon Chamber (MUC) consisting of 1000 $\mathrm{m}^2$ of resistive plate chambers is used to identify muons and it provides a spatial resolution better than 2 cm. More information can be found in~\cite{bes3}.\\
BESIII has accumulated the world$^\prime$s largest data samples of $e^+e^−$ collisions in the $\tau$-charm region. In the next sections the data samples important for the measurement of baryon form factors presented here will be highlighted.

\section{Measurement of $e^+e^- \rightarrow p \overline{p}$ and $e^+e^- \rightarrow p \overline{p} \gamma$}
While so far a number of experiments have measured the cross section of the channel $e^+e^- \rightarrow p\bar{p}$~\cite{bes,CLEO,BABARpp,PS170,E760,E835,cmd3}, almost no measurements exist concerning the electromagnetic form factors of the proton~\cite{BABARpp,PS170,cmd3}. Furthermore, close to the $p\bar{p}$ production threshold the measurements by BaBar~\cite{BABARpp} and PS170~\cite{PS170} show strong tension.

\begin{figure}[t!]
\begin{center}
\begin{minipage}{0.45\linewidth}
\centerline{\includegraphics[width=1\linewidth]{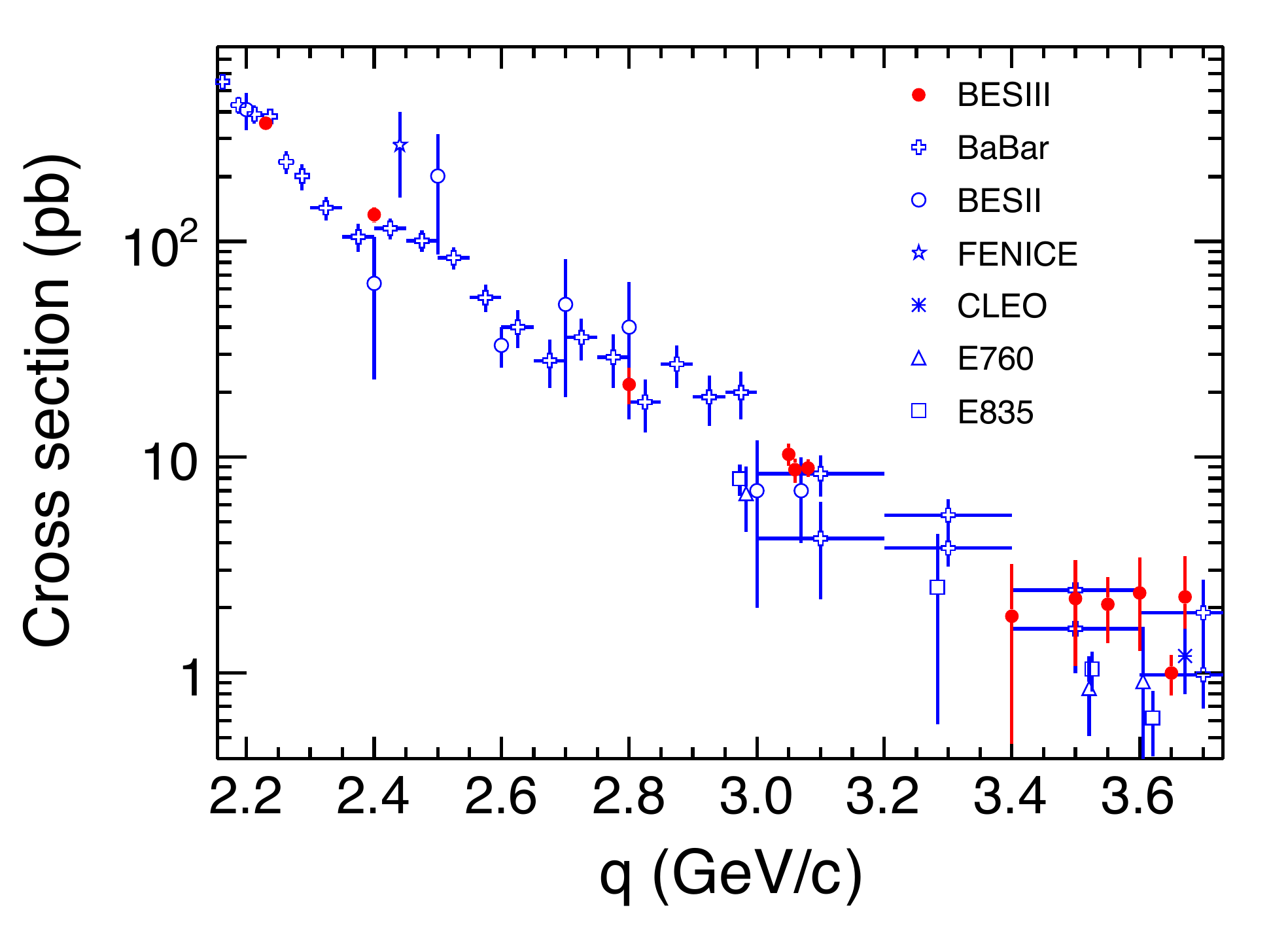} }
\end{minipage}
\begin{minipage}{0.435\linewidth}
\centerline{\includegraphics[width=1\linewidth]{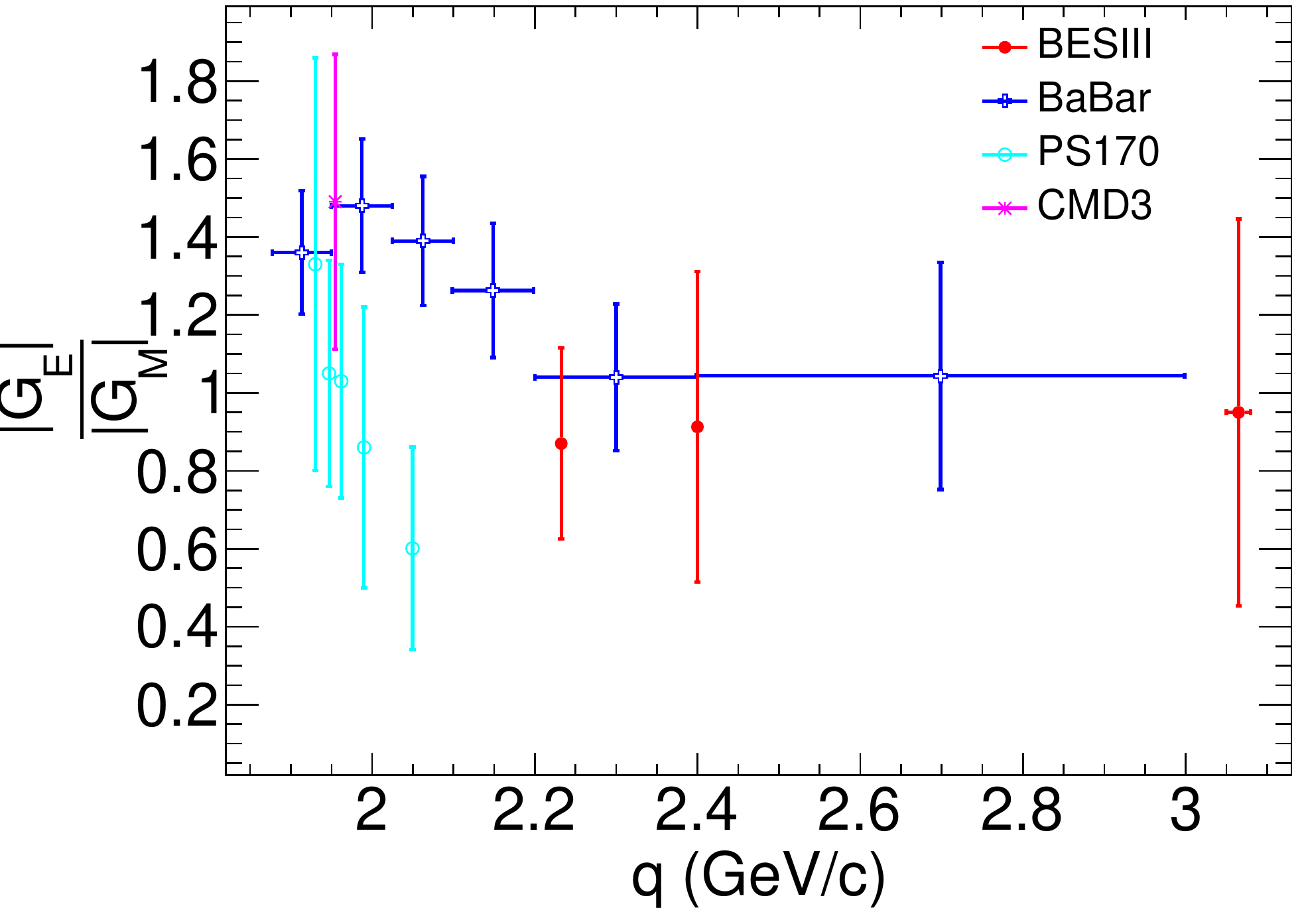}}
\end{minipage}
\end{center}
\caption[]{(Left) Measurements of the Born cross section of $e^+e^- \to p \bar{p}$ (left) and $|G_E/G_M|$ (right) in the time-like region. BESIII results are shown in red~\cite{xiaorong}.
}
\label{FFxiaorong}
\end{figure}

BESIII has published the measurement of the channel $e^+e^-\rightarrow p\bar{p}$ at 12 center-of-mass energies between 2.2324 and 3.6710 GeV~\cite{xiaorong}. These data were collected in 2011 and 2012 and correspond to a luminosity of 157 $\mathrm{pb^{-1}}$. The Born cross section was extracted according to 
\begin{equation}
\sigma_\mathrm{Born} =  \frac{N_{obs} - N_{bkg}}{\mathcal{L}\cdot\epsilon(1+\delta)}, 
\label{xsmeasured}
\end{equation}
where the number of background events is subtracted from the observed signal event candidates, normalized with the luminosity at each $\sqrt{s}$, $\mathcal{L}$, and corrected with the selection efficiency of the process, $\epsilon$,  times the ISR radiative correction factor up to next-to-leading order (NLO), $(1+\delta)$. The ConExc generator~\cite{conexc} was used both for the efficiency and the radiative factor evaluation.
The accuracy in the cross section measurements was between 6.0$\%$ and 18.9$\%$ up to $\sqrt{s} < 3.08$ GeV and the results are shown in Fig.~\ref{FFxiaorong} (left) together with previous experimental results~\cite{bes,CLEO,BABARpp,PS170,E760,E835}.  The EFF was extracted according to Eq.~\ref{effectiveff}. A fit to the polar angular distribution of the proton in c.m. was performed according to Eq.~\ref{diff} and $|G_E/G_M|$ and $|G_M|$ were extracted for three $\sqrt{s}$ (Fig.~\ref{FFxiaorong} (right)). Table~\ref{tab_ppbar1} summarizes the results. While the measurements by the different experiments concerning the EFF show very good agreement, this is not the case for $|G_E/G_M|$, where the measurements by BaBar~\cite{BABARpp} and PS170~\cite{PS170} disagree for low momentum transfer. 
\begin{table}[h!]
\centering
\caption{Results for $|G_E/G_M|$ and $|G_M|$ obtained by BESIII.}
\label{tab_ppbar1}
\tabcolsep7pt
\begin{tabular}{lccc}
\hline
$\sqrt{s}$~(GeV)  &$|G_E/G_M|$   & $|G_M|$~($\times10^{-2}$)  \\
\hline
2.2324  &$0.87\pm0.24\pm0.05$  &$18.42\pm5.09\pm0.98$  \\
2.4000  &$0.91\pm0.38\pm0.12$  &$11.30\pm4.73\pm1.53$  \\
(3.0500-3.0800) &$0.95\pm0.45\pm0.21$ &$3.61\pm1.71\pm0.82$ \\
\hline
\end{tabular}
\end{table}

\begin{figure}[t!]
\begin{center}
\begin{minipage}{0.4\linewidth}
\centerline{\includegraphics[width=1\linewidth]{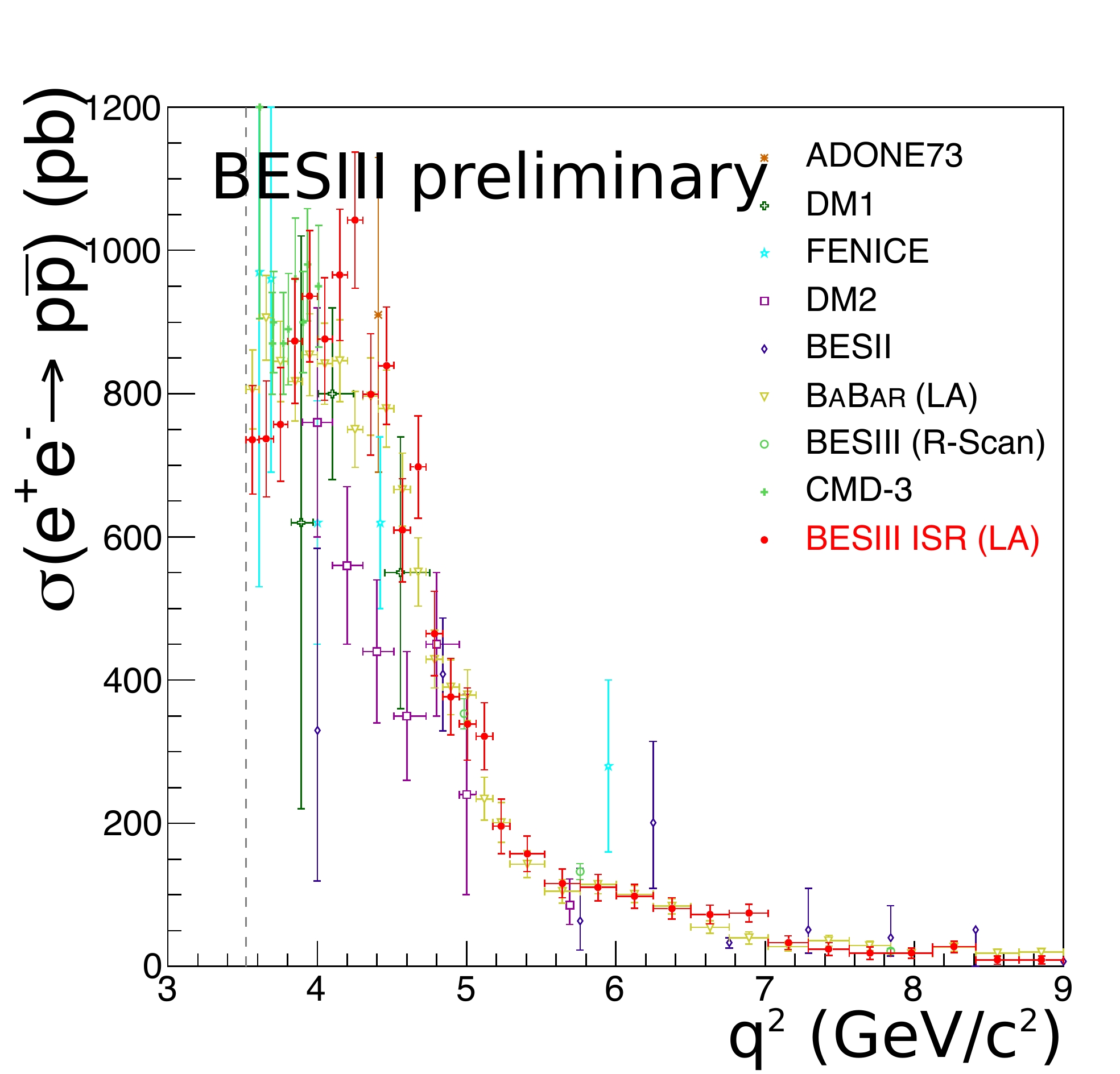}}
\end{minipage}
\begin{minipage}{0.4\linewidth}
\centerline{\includegraphics[width=1\linewidth]{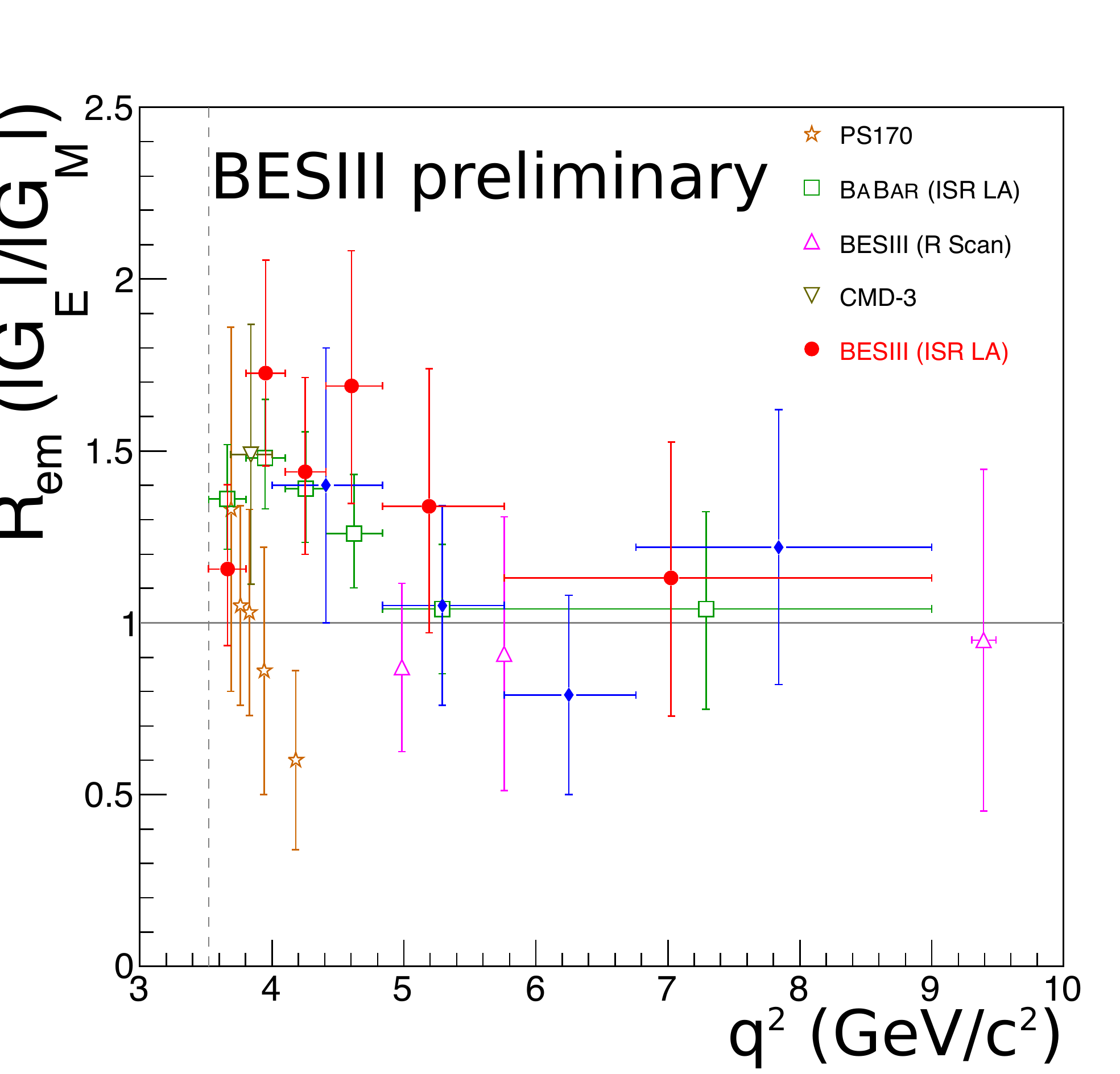}}
\end{minipage}
\end{center}
\caption[]{Measurements of the cross section of $e^+e^- \to p \bar{p}$ (left) and $|G_E/G_M|$ (right) in the time-like region. Preliminary results from BESIII from the study of $e^+e^- \to p \bar{p} \gamma$ with the ISR photon tagged are shown in red.}
\label{isrproton}
\end{figure}

At present, the precision in the measurement of $|G_E/G_M|$ of the proton is dominated by statistics. From January till June 2015, BESIII performed a high luminosity low-energy scan with about 650 $\mathrm{pb^{-1}}$ of luminosity collected at 22 different center-of-mass energies between 2.0 and 3.08 GeV (black dots in Fig.~\ref{lambdac} (left)). From the analysis of these data, it is expected the measurement of $|G_E/G_M|$ and $|G_M|$ with statistical accuracies below the $10\%$ level and at the level of few percent, respectively. \\
The study of the $e^+e^{-} \rightarrow p \bar{p}$ process is also possible from the measurement of $e^+e^{-} \rightarrow p \bar{p} \gamma$ events where the photon is emitted from the initial state. A total statistics of 7.41 $\mathrm{fb}^{-1}$ collected at the main charmonium resonances and Y-states has been used for these studies.
Two kinds of scenarios can be distinguished depending on whether the radiated photon can be detected (tagged case, about 12$\%$ of the events) or not (untagged case, about $47\%$ of the events). 
BESIII has preliminary results from the analysis of the tagged photon events. The ratio $|G_E/G_M|$ has been extracted for six different bins of momentum transfer starting from the proton-antiproton production threshold and up to 3.0 GeV/$c$ (Fig.~\ref{isrproton} (right)). 
The statistical and systematic uncertainties range between 18.5$\%$ to 33.6$\%$ and 4.5$\%$ to 15.6$\%$, respectively. The Born cross section of the process  $e^+e^{-} \rightarrow p \bar{p}$ and the proton efffective form factor have also been measured and the results are in good agreement with the existing measurements (Fig.~\ref{isrproton} (left)). 
From the analysis of the untagged photon events, competitive results compared to previous measurements are expected for the cross section and the proton effective form factor from 2.0 to 3.8 GeV/$c$. However, due to the angular distribution of these events, no significant improvement is expected for  $|G_E/G_M|$. 

The study of the process $e^+e^- \rightarrow p \bar{p}$  with two different techinques, energy scan and initial state radiation, allows complementary approaches to the measurement of $\sigma(e^+e^- \to p \bar{p})$ and $|G_E/G_M|$. While the energy scan technique provides high accuracy measurements at well defined energies, the use of the ISR technique allows a continuous measurement over a wide region of momentum transfer. Thus the energy scan technique, is essential for the measurement of FFs, while the ISR technique is very useful to study the presence of structures in the hadronic cross section. Indeed, the steps first observed in  BaBar's data in the energy region close to the $p\bar{p}$ production threshold and up to the J/$\Psi$ peak~\cite{BABARpp}, have recently triggered different theoretical interpretations~\cite{egle,meissner}. Another interesting field of theoretical investigations is the behavior of the $e^+e^- \to p \bar{p}$ cross section very close to the $p\bar{p}$ production threshold~\cite{ppthreshold1,ppthreshold3}. The final state interaction of the $p\bar{p}$ sytem, enhances the cross section near the production threshold as compared to the phase space, and is responsible for the peak and the rapid fall near threshold. 

 

\section{Measurement of $e^+e^- \rightarrow \Lambda \overline{\Lambda}$ and prospects for hyperon form factors}
The structure of hyperons is a rather unexplored territory~\cite{LAMBDAS2,LAMBDAS,Dobbs}. Since hyperons are unstable, they are difficult to study in the space-like region - hyperon targets are unfeasible and the quality of hyperon beams is in general not sufficient. Time-like form factors therefore offer the best possibility to study hyperon structure. 

\begin{figure}[t!]
\begin{center}
\begin{minipage}{0.45\linewidth}
\centerline{\includegraphics[width=1\linewidth]{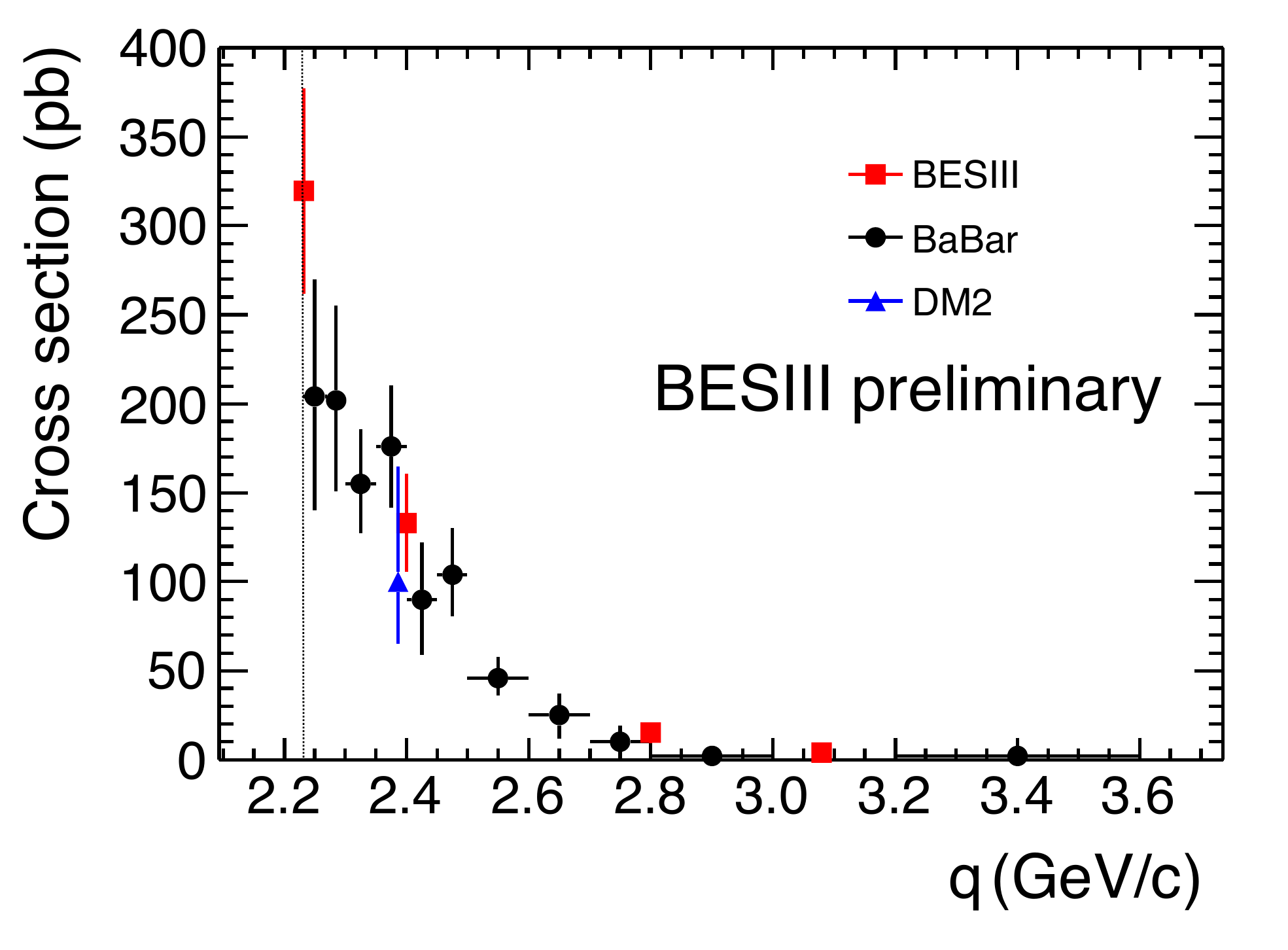}}
\end{minipage}
\begin{minipage}{0.45\linewidth}
\centerline{\includegraphics[width=1\linewidth]{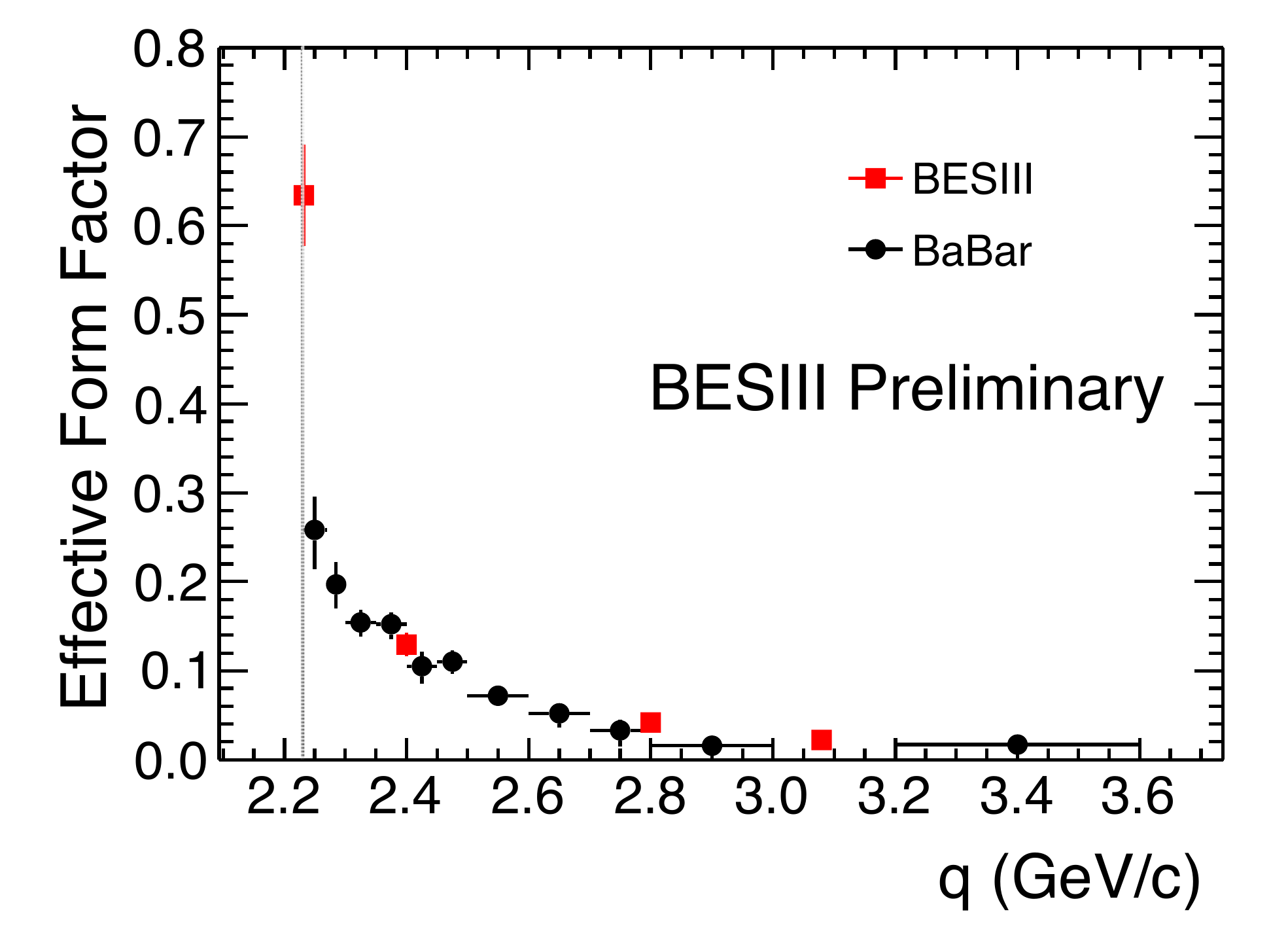}}
\end{minipage}
\end{center}
\caption[]{(Left) Measurements of $e^+e^{-} \rightarrow \Lambda \bar{\Lambda}$ cross section and (right) $\Lambda$ effective form factor.}
\label{lambda}
\end{figure}

BESIII has preliminary results on the measurement of the channel $e^+e^{-} \rightarrow \Lambda \bar{\Lambda}$. The analysis was based on $40.5~\mathrm{pb}^{-1}$ collected at four different energy scan points during 2011 and 2012. The lowest energy point is  2.2324 GeV, only 1 MeV above the $\Lambda\bar{\Lambda}$-threshold. This makes it possible to measure the Born cross section almost at threshold. To use as much statistics as possible, both events where $\Lambda$ and $\bar{\Lambda}$ decay to the charged mode ($\mathrm{BR}(\Lambda \rightarrow p\pi^-) = 64\%$) and events where the $\bar{\Lambda}$ decays to the neutral mode  ($\mathrm{BR}(\bar{\Lambda} \rightarrow \bar{n} \pi^0) = 36\%$) were selected. In the first case, the identification relied on finding two mono-energetic charged pions and a possible $\bar{p}$-annihilation.
In the second case, the $\bar{n}$-annihilation was identified through the use of multi-variate analysis of variables provided by the electromagnetic calorimeter. Additonally, a mono-energetic $\pi^0$ was reconstructed to fully identify the channel. 
For the higher energy points, only the charged decay modes of $\Lambda$ and $\bar{\Lambda}$ were reconstructed by identifying all the charged tracks and using the event kinematics. The preliminary results on the measurement of the Born cross section are shown in Figure~\ref{lambda} (left) together with previous measurements~\cite{LAMBDAS,LAMBDAS2}. The Born cross section at threshold is found to be $318 \pm 47 \pm 37$ pb. Given that the Coulomb factor in Eq.~\ref{diff} is equal to 1 for neutral baryon pairs, the cross section is expected to go to zero at threshold. This result confirms BaBar$^\prime$s measurement~\cite{LAMBDAS2} but with much higher accuracy in the momentum transfer. 
The observed threshold enhancement implies that a more complicated physics scenario is underlying. The unexpected features of baryon pair production near threshold have driven a lot of theoretical studies~\cite{llbarthreshold1}, including scenarios that invoke bound states or unobserved meson resonances. It was also interpreted as an attractive Coulomb interaction on the constituent quark level~\cite{ppthreshold3}.
The BESIII measurement improves previous results at low momentum transfer at least by 10$\%$ and even more above 2.4 GeV/c. The lambda EFF extracted from the cross section measurement is shown in Fig.~\ref{lambda} (right).

Baryon form factor measurements are one of the most important reasons why BESIII has collected an unprecedented amount of off-resonance data. As mentioned before, the major part was collected in 2015 within the the center-of-mass energy range of 2.0 to 3.08 GeV, and in 2014 above the production threshold of the $\Lambda_c\overline{\Lambda}_c$ pair. Figure~\ref{lambdac} (left) shows the integrated luminosites collected at different energies by BESIII and by other experiments located in $e^+e^-$ colliders. The production thresholds of different channels are also shown. All baryons in the ground state spin 1/2 SU(3) octet and spin 3/2 decuplet are accessible using BESIII scan data.  From the analysis of the 2015 data, it is expected that the ratio of the absolute values of the lambda electromagnetic form factors, $|G_E/G_M|$, can be measured at five energy points. Furthermore, at 2.396 GeV, the data sample is expected to be sufficient for the measurement of the phase, $\Delta\Phi$, allowing the full determination of the lambda electromagnetic form factors for the first time. Concerning the $\Sigma$ hyperon, it is expected that the effective form factors of the triplet can be determined in several points, as well as the ratio of the $\Sigma^+$ electromagnetic form factors. In the case of multi-strange hyperons, it is expected that the statistics are enough to measure their effective form factors at several energies.

\section{Measurement of $e^+e^- \rightarrow \Lambda_c \overline{\Lambda}_c$ close to the production threshold}

\begin{figure}[t!]
\begin{center}
\begin{minipage}{0.435\linewidth}
\centerline{\includegraphics[width=1\linewidth,height=0.75\linewidth]{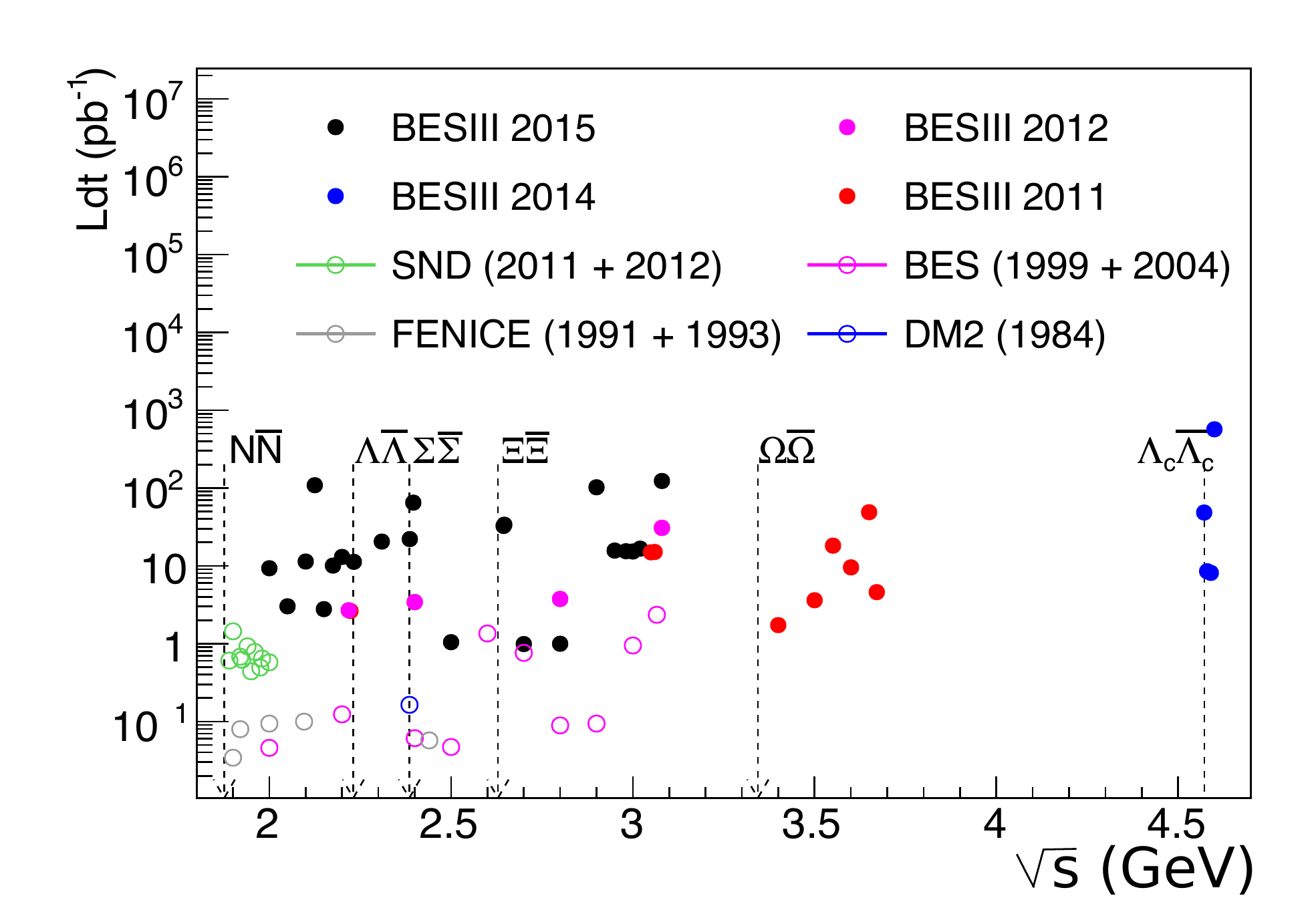}}
\end{minipage}
\begin{minipage}{0.45\linewidth} \vspace{0.5cm}
\centerline{\includegraphics[width=1\linewidth,height=0.76\linewidth]{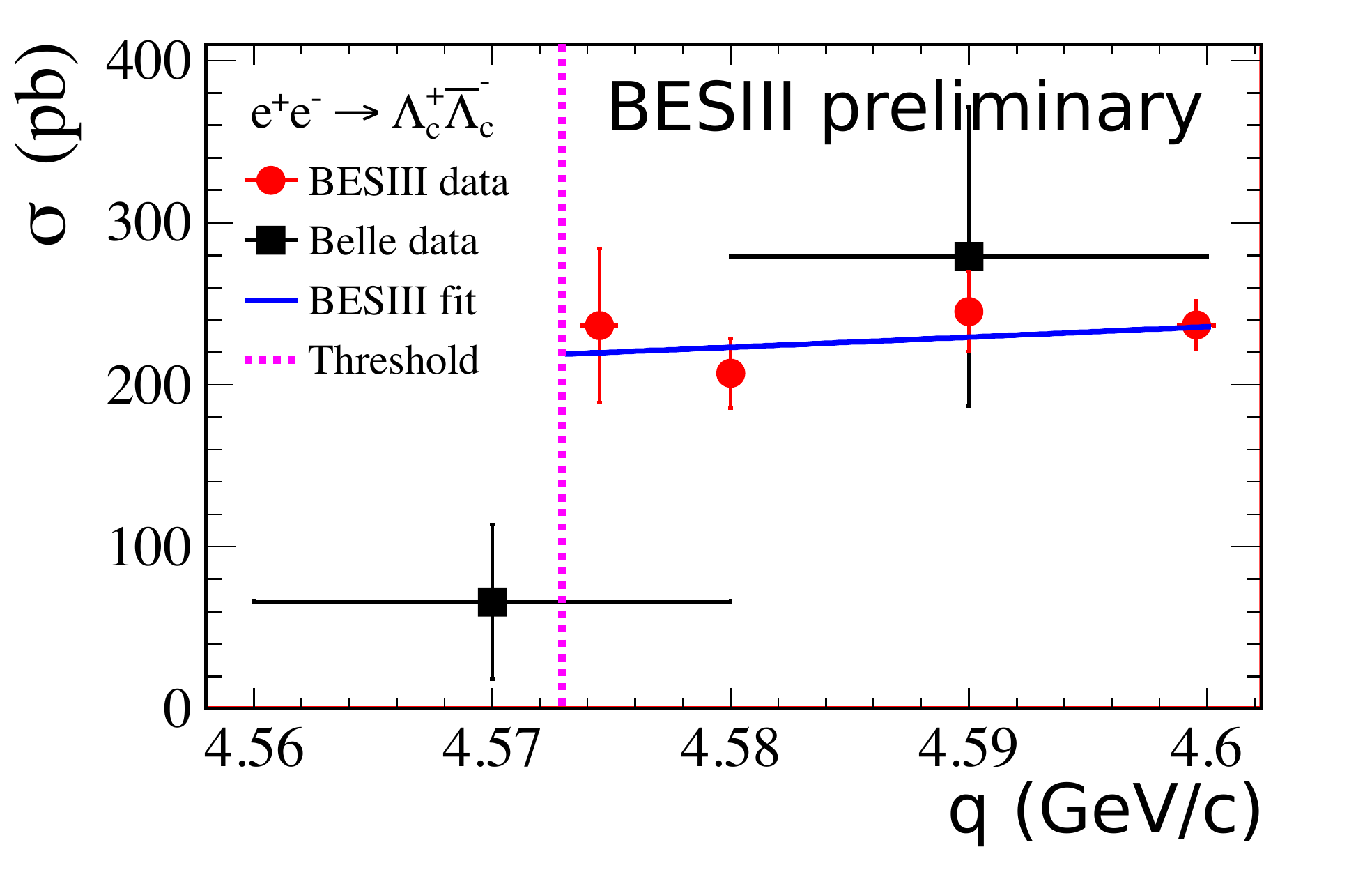}}
\end{minipage}
\end{center}
\caption[]{(Left) The integrated luminosity collected by BESIII during different energy scan periods outside resonances. Shown are also the luminosities collected by other experiments at different $e^+e^-$ colliders.Thresholds of important baryon-antybaryon production channels are marked with vertical dashed lines. (Right) The Born cross section of $e^+e^- \to \Lambda_c^+ \bar{\Lambda}_c^-$ obtained by BESIII and Belle~\cite{belle}. BESIII results are preliminary. The blue solid curve represents the input line-shape in the generator to determine $f_{ISR}$.}
\label{lambdac}
\end{figure}

BESIII has preliminary results on the measurement of the channel  $e^+e^- \to \Lambda_c^+ \bar{\Lambda}_c^-$. The analysis was based on $631.3~\mathrm{pb}^{-1}$ collected at four different energy scan points in 2014, $\sqrt{s} = $4.5745, 4.5809, 4.5900 and 4.5995 GeV. The lowest energy point is only 1.6 MeV above the $\Lambda_c^+ \bar{\Lambda}_c^-$ threshold. At each of the energy points, ten Cabibbo-favored hadronic decay modes, $\Lambda_c^+ \rightarrow p K^-\pi^+$ , $p K_S^0$ , $\Lambda \pi^+$ , $p K^- \pi^+ \pi^0$ , $p K^0 \pi^0$ , $\Lambda \pi^+ \pi^0$ , $p K_S \pi^+ \pi^-$ , $\Lambda \pi^+ \pi^+ \pi^-$ , $\Sigma^0 \pi^+$ , and $\Sigma^+\pi^+\pi^-$, as well as the ten corresponding charge-conjugate modes are independently single tagged. The Born cross section of the $i$-th mode is calculated as:
\begin{equation}
\sigma_i = \frac{N_i}{\epsilon_i \cdot \mathcal{L} \cdot f_{VP} \cdot BR_i \cdot f_{ISR}},
\end{equation}
where $N_i$ and $\epsilon_i$ represent the yield and the detection efficiency, respectively, $\mathcal{L}$ stands for the integrated luminosity, $f_{VP}$ is the vacuum polarization correction factor, $BR_i$ represents the products of BRs of the $i$-th $\Lambda_c$ decay mode and its subsequent decay(s), and $f_{ISR}$ is the ISR correction factor calculated iteratively. The total Born cross section is obtained from the weighted average over the 20 individual measurements 
$\sigma = \sum_i w_i\sigma_i $,
with $w_i$ the weight of the $i$-th cross section. The results are shown in Fig.~\ref{lambdac} (right). A rise of the cross-section just above threshold much steeper than the phase-space expectation is discerned. 

\begin{figure}[t]
\begin{center}
\begin{minipage}{0.45\linewidth}
\centerline{\includegraphics[width=1\linewidth,height=0.75\linewidth]{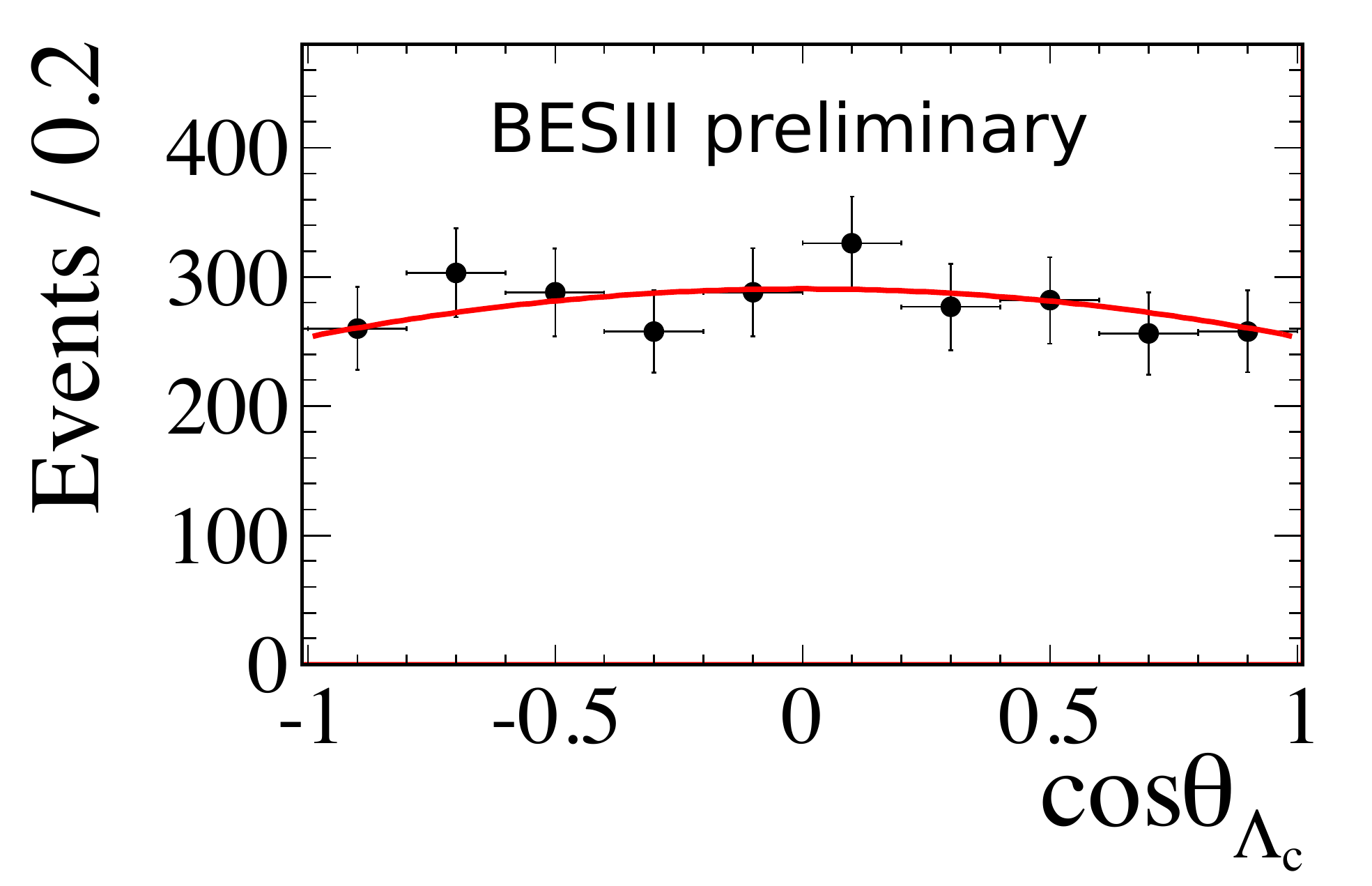}}
\end{minipage}
\begin{minipage}{0.45\linewidth}
\centerline{\includegraphics[width=1\linewidth,height=0.75\linewidth]{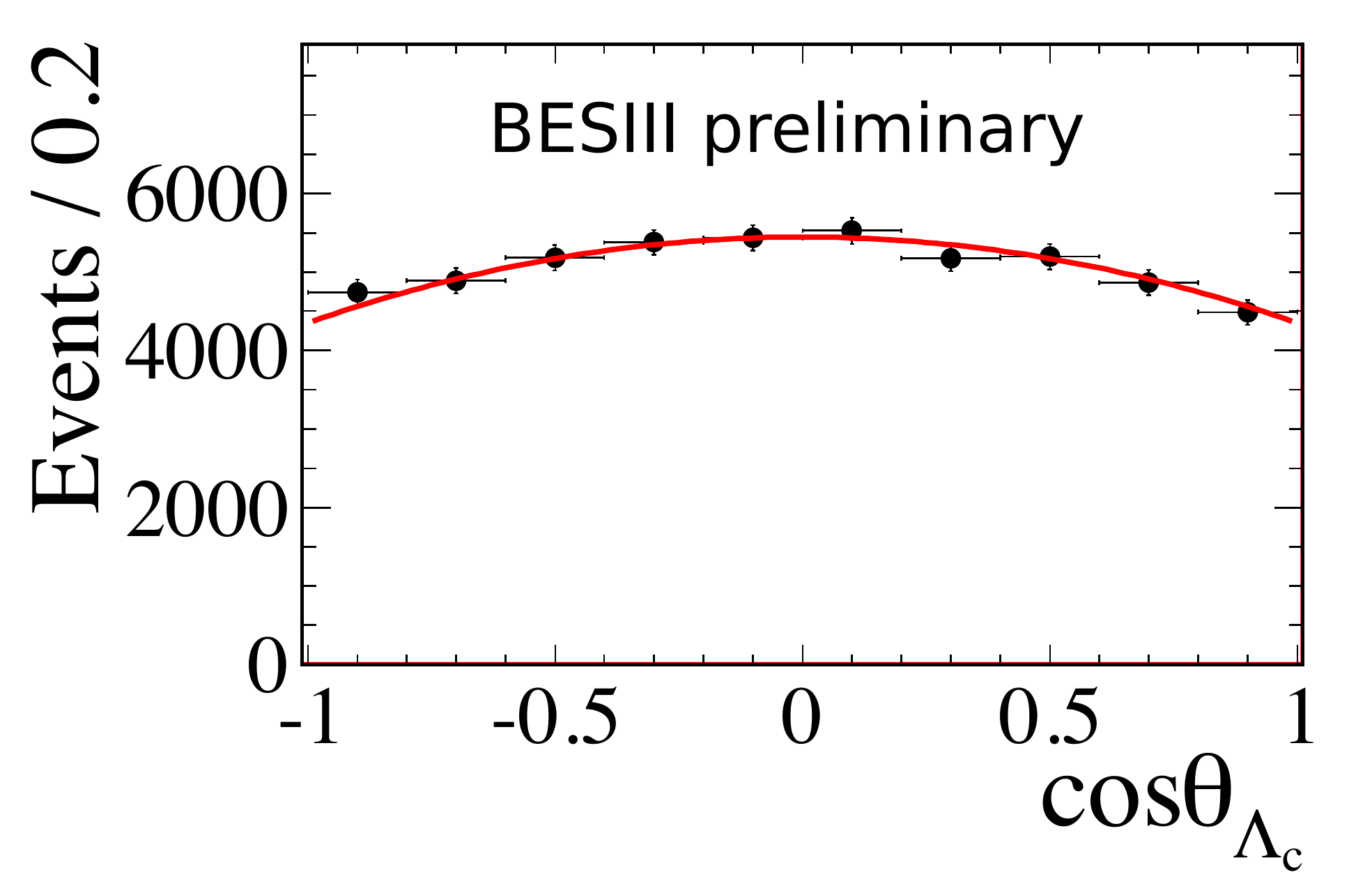}}
\end{minipage}
\end{center}
\caption[]{The angular distribution and corresponding fit results in data at $\sqrt{s} = 4.5745$ GeV (left) and $\sqrt{s} = 4.5995 $GeV (right) .}
\label{lambdac2}
\end{figure}

The higher statistic data samples at $\sqrt{s} = $ 4.5745 and 4.5995 GeV enable the study of the polar angular distribution of $\Lambda_c$ in the $e^+e^-$ center-of-mass system. From these distributions, the ratios between the electric and magnetic form factors, $|G_E/G_M|$, have been extracted for the first time. The shape function $f(\theta) \propto  (1 + \alpha_{\Lambda_c}cos^2\theta)$ is fitted to the combined data contaning the yields of $\Lambda_c^+$ and $\bar{\Lambda}_c^-$ to all the 10 decay modes (Fig.~\ref{lambdac2}). The $|G_E/G_M|$ ratios are then extracted using:
\begin{equation}
|G_E/G_M|^2(1-\beta^2) = (1 - \alpha_{\Lambda_c})/ (1 + \alpha_{\Lambda_c})
\end{equation}
and are shown in Table~\ref{tab_lambda}.

\begin{table}[h!]
\begin{center}
\caption{Shape parameters of the angular distribution and $|G_E/G_M|$ ratios at  $\sqrt{s} = 4.5745$ GeV and $\sqrt{s} = 4.5995 $GeV. The uncertainties are statistical and systematic, respectively.}
\label{tab_lambda}
\tabcolsep7pt\begin{tabular}{lccc}
\hline
$\sqrt{s}$~(GeV)  &$\alpha_{\Lambda_c}$   & $|G_E/G_M|$  \\
\hline
4.5745  &$-0.13\pm0.12\pm0.08$  &$1.14\pm0.14\pm0.07$  \\
4.5995  &$-0.20\pm0.04\pm0.02$  &$1.23\pm0.05\pm0.03$  \\
\hline
\end{tabular}
\end{center}
\end{table}

In a future extension of this analysis, also polarization observables like the polarization, $P_n$,  and the spin correlation of the outgoing $\Lambda_c$ pair, $C_{nl}$, will be extracted, making possible the full determination of the $\Lambda_c$ electromagnetic form factors.

\section{Summary}
Electromagnetic form factors provide a quantitative description of hadron structure and are basic observables of QCD. Baryon form factors can be studied in the time-like region by the BESIII experiment using the scan and the initial-state radiation techniques. A pilot energy scan was performed in 2011-2012, resulting in new measurements on proton and $\Lambda$ hyperon electromagnetic form factors. Also preliminary results exist for the $\Lambda_c$ electromagnetic form factors close to the $\Lambda_c^+ \bar{\Lambda}_c^-$ production threshold. A  world leading data sample collected in 2015 will enable precison measurements of proton electromagnetic form factors, the full determination of the $\Lambda$ form factors, and first measurements of heavier single and multistrange hyperons effective form factors. Furthermore, also neutron eletromagnetic form factors will be measured with unprecedented precision in the time-like region. BESIII has also preliminary results on the $e^+e^-\to p \bar{p}$ channel using the ISR technique. The results complement our knowledge of the cross section of the process and the ratio of the proton eletromagnetic form factors in a wide and continuous region of momentum transfer starting from the $p\bar{p}$ production threshold. More results are coming soon.

\section*{Acknowledgments}
This work was supported  by the German Research Foundation DFG under the Collaborative Research Center CRC-1044.

\section*{References}


\end{document}